\documentclass[12pt,preprint]{aastex}

\shorttitle{TRAVEL-TIME INHOMOGENEITIES IN SUNSPOTS }
\shortauthors{MORADI, HANASOGE \& CALLY}
\begin{document}

\title{Numerical models of travel-time inhomogeneities in sunspots}
\author{H. Moradi}
\affil{Centre for Stellar and Planetary Astrophysics, School of Mathematical Sciences, Monash University, Victoria 3800, Australia}
\email{hamed.moradi@sci.monash.edu.au}    
\author{S. M. Hanasoge\altaffilmark{1}}
\affil{W. W. Hansen Experimental Physics Laboratory, Stanford University, Stanford, CA 94305, U.S.A.}
\altaffiltext{1}{Visiting Scientist: Centre for Stellar and Planetary Astrophysics, School of Mathematical Sciences, Monash University, Victoria 3800, Australia}    
\author{P. S. Cally}
\affil{Centre for Stellar and Planetary Astrophysics, School of Mathematical Sciences, Monash University, Victoria 3800, Australia}

\begin{abstract}
We investigate the direct contribution of strong, sunspot-like magnetic fields to helioseismic wave travel-time shifts
via two numerical forward models, a 3D ideal MHD solver and MHD ray theory. The simulated data cubes are analyzed using the traditional time-distance center-to-annulus measurement technique. We also isolate and analyze the direct contribution from purely thermal perturbations to the observed travel-time shifts, confirming some existing ideas and bring forth new ones: (i) that the observed travel-time shifts in the vicinity of sunspots are largely governed by MHD physics, (ii) the travel-time shifts are sensitively dependent on frequency and phase-speed filter parameters and the background power below the $p_1$ ridge, 
and finally, (iii) despite its seeming limitations, ray theory succeeds in capturing the essence of the travel-time variations as derived from the MHD simulations.  
\end{abstract}

\section{INTRODUCTION}
Local helioseismic diagnostic methods such as time-distance helioseismology \citep{duvalletal1993}, helioseismic holography \citep{lb1997} and ring-diagram analysis \citep{hill1988}, have over the years provided us with unprecedented views of the structures and flows under sunspots and active regions. However, a growing body of evidence appears to suggest that interpretations of the measured statistical changes in the properties of the wave-field may be rendered inaccurate by complexities associated with the observations and wave propagation physics. Incorporating the full MHD physics and understanding the contributions of phase and frequency filters, and differences in the line formation height, are thought to be central to future models of sunspots.

One of the earliest studies that highlighted the interaction of waves with sunspots was the Fourier-Hankel analysis of \cite{bdl1987}, who found that sunspots can absorb up to half of the incident acoustic-wave power and shift the phases of interacting waves quite significantly (see also \citealt{braun1995}). These results were echoed over the years by a steady steam of theoretical results (e.g. \citealp{bogdanetal1996}; \citealp{cb1997}; \citealp{ccb2003}; \citealp{crouchetal2005}; \citealp{cally2007}) that have consistently emphasized the need for more sophisticated modeling and interpretation of wave propagation in strongly magnetized regions. 

Important advances in our observational understanding of sunspots were also achieved by \cite{duvalletal1996} and \cite{zkd2001}, who inferred the presence of flows underneath sunspots, and \cite{kds2000} who estimated the sub-surface wave-speed topology. However, while the inversion procedures applied to derive these results fail to directly account for the tensorial nature of magnetic field effects, the action of the field is mimicked via changes in the acoustic properties of the medium (the so-called wave speed). Recently however, numerical forward models of helioseismic wave (e.g. \citealp{cgd2008}; \citealp{hanasoge2008}) and ray \citep{mc2008} propagation in magnetized atmospheres have been developed and are beginning to make inroads into this problem. In particular, the results of \cite{mc2008} and Cameron (2008; private communication) strongly suggest that active-region magnetic fields play a substantial role in influencing the wave field, and that the complex interaction of magnetic fields with solar oscillations, as opposed changes in the wave-speed, are the major causes of observed travel-time inhomogeneities in sunspots. 

We study the impact of strong magnetic fields on wave propagation and the consequences for time-distance helioseismology using two numerical forward models, a 3D ideal MHD solver and MHD ray theory. The simulated data cubes are analyzed using the traditional surface-focused center-to-annulus method frequently applied in the time-distance analyses of sunspots (e.g., \citealp{cbk2006}). Furthermore, we apply the same method as \cite{mc2008} to also isolate and analyze the thermal contribution to the observed travel-time shifts.  
 
\section{THE SUNSPOT MODEL}\label{sunspot}
\begin{figure}[ht]
\begin{center}
\epsscale{1.0} 
\plotone{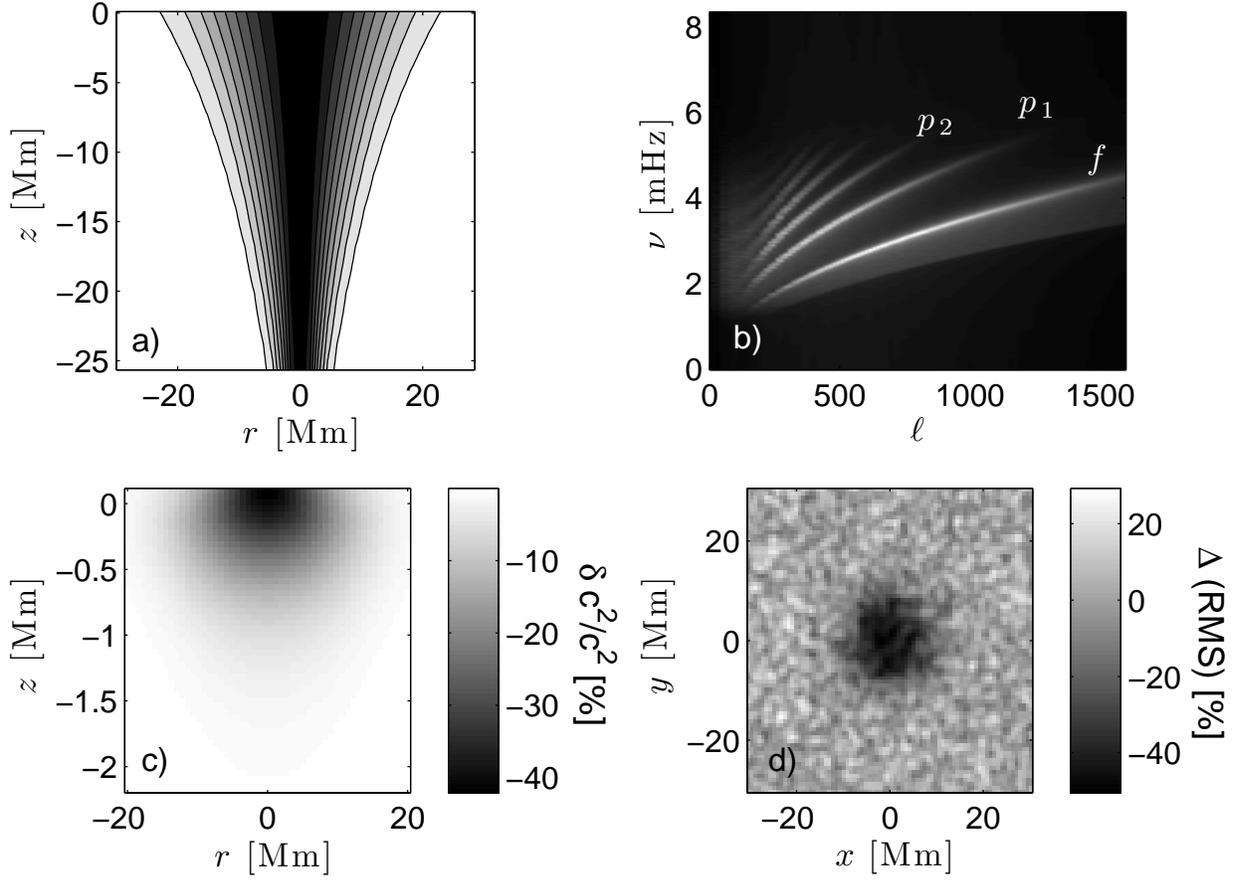}
\end{center}
\caption{Some properties of the model atmosphere: a) shows the field configuration of the sunspot model, depicted as lines of constant magnetic flux, b) represents the power spectrum of the simulated Doppler velocity data-cube with the location of the $p$- and $f$-mode ridges, c) depicts the near-surface sound speed/thermal profile and d) shows a power map normalized to the quiet Sun. }
\label{fig:mhd}
\end{figure}

The background stratification is given by a convectively stable, hydrostatic $m=2.15$ polytrope below the photosphere, smoothly connected to an isothermal atmosphere. The magneto-hydrostatic (MHS) sunspot model that we embed in the background is similar in construction to that of \cite{cgd2008} and \cite{hanasoge2008}, where the flux tube is modeled by an axisymmetric magnetic field geometry based on the \cite{st1958} self-similar solution. This approximation requires the following choices for the radial ($B_r$) and vertical ($B_z$) components of the magnetic field: 
\begin{equation}
B_z=M\psi(z)e^{-r^2\psi(z)},\,
B_r=-M\frac{r}{2}\frac{d\psi}{dz} e^{-r^2\psi(z)},
\end{equation}
where  $r$ and $z$ are the radial and vertical coordinates  in cylindrical geometry. The term $M$ controls the magnitude of the magnetic field (peak field strength of 3000~G at  the photosphere) and the flux ($\pi M$). The horizontal extent of the tube and its rate of divergence with altitude is set by $\psi(z)$ (see Figure~\ref{fig:mhd} a). 

Upon solving the MHS equations of pressure and Lorentz support (described in detail in \citealp{cgd2008, hanasoge2008, mc2008}), we obtain the altered thermodynamic stratification of the underlying magnetized plasma (Figure~\ref{fig:mhd} c). 

\section{FORWARD MODELS}\label{forward}
\subsection{MHD Wave-Field Simulations} \label{MHD.hanasoge}
The linearized ideal MHD wave equations are integrated according to the recipe of \cite{hanasoge2008}.
We implement periodic horizontal boundaries and place damping sponges adjacent to the vertical boundaries
to enhance the absorption and transmission of outgoing waves. Waves are excited via a pre-computed
deterministic source function that acts on the vertical momentum equation. In a manner similar to \cite{hanasogeetal2007}, the forcing term is also multiplied by a spatial function that mutes source activity in a circular region of 10 Mm radius to simulate the suppression of granulation related wave sources in a sunspot. 
Figure~\ref{fig:mhd} shows some of the resulting properties of the simulated vertical (Doppler) velocity data extracted from the MHD simulations. 

\subsection{MHD Ray-Path Simulations}\label{MHD.moradi}
\cite{mc2008} detail the steps involved in using MHD ray theory to model helioseismic ray propagation in magnetized atmospheres. 
Here, we provide a brief description of the magneto-acoustic ray tracing procedure.  

The ray paths are computed in Cartesian geometry in the vertical $x$-$z$ plane assumed to contain both magnetic field lines and ray paths. In this case, we only require the 2D dispersion relation with the Alfv\'en wave factored out: 
\begin{eqnarray}
\mathcal{D} = \omega^4 - (a^2+c^2) \omega^2K^2 + a^2c^2K^2k^2_\parallel + c^2N^2k^2_x - (\omega^2-a^2_zK^2)\omega^2_c=0
\label{eq:disp}
\end{eqnarray}
where $K=|\textbf{k}|$, $c$ represents the sound speed, $a$ the Alfven speed, $N^2$ is the squared Brunt-V\"ais\"al\"a  frequency and $\omega^2_c$ is the square of the isothermal acoustic cut-off frequency. The remaining term, $k_\parallel=\hat\textbf{B}_0.\textbf{k}$ (where $\textbf{B}_0$ is the prescribed magnetic field), represents the component of the wavevector $\bf{k}$ parallel to the magnetic field. The full 3D dispersion relation is presented in \cite{mc2008}. The construction of $\bf{k}$ is completed by specifying the governing equations of the ray paths using the zeroth order eikonal approximation \citep{weinberg1962,mc2008} which are solved using a fourth-order Runge-Kutta numerical method. The magneto-acoustic rays stay on the fast-wave dispersion branch at all times. It should be noted that neither forward model (i.e., $\S$\ref{MHD.hanasoge}, \ref{MHD.moradi}) accounts for the presence of flows.

\section{Travel-Time Measurement Scheme}
\begin{figure}[ht]
\epsscale{1.0}
\begin{center} 
\plotone{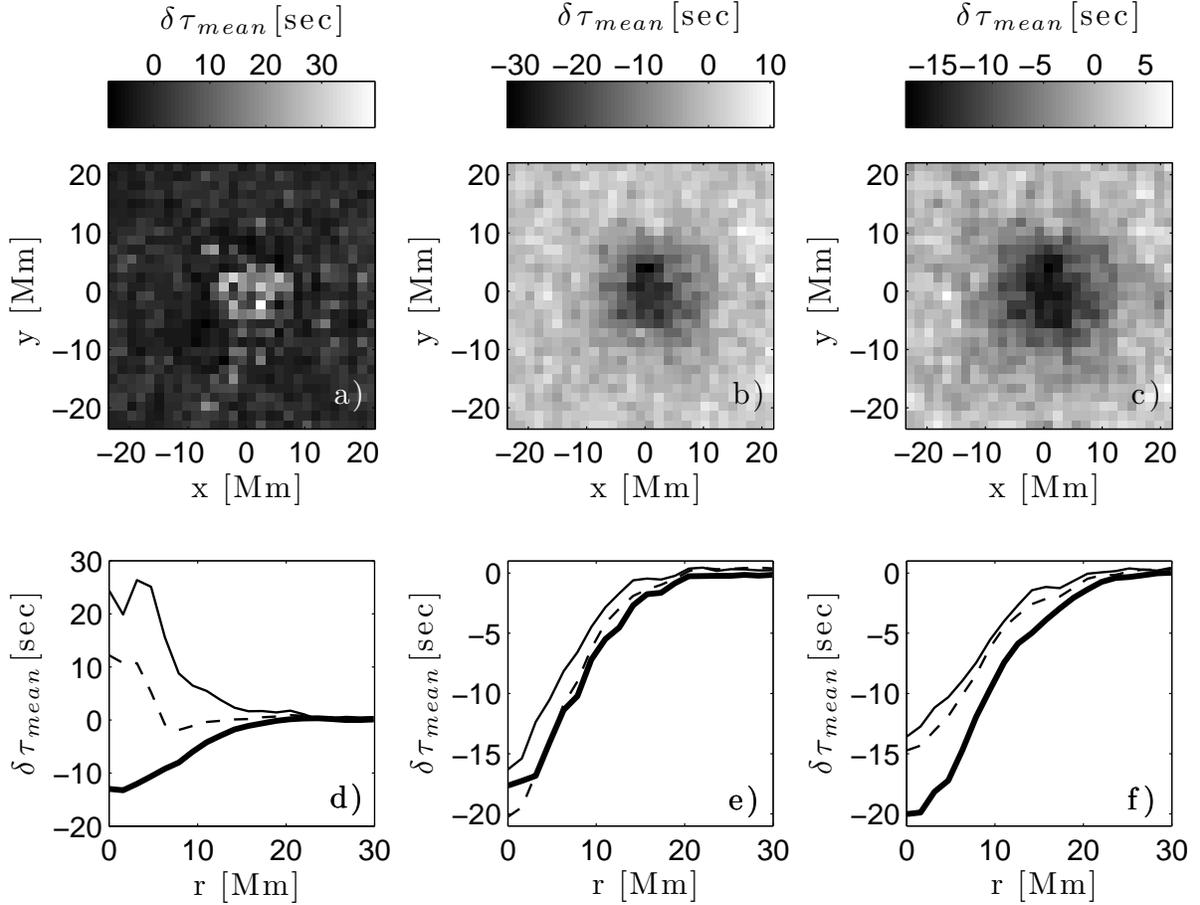}
\end{center}
\caption{Center-to-annulus time-distance $\delta\tau_{mean}$ maps (no frequency filtering) and azimuthal averages for $\Delta$= 3.7-8.7 (a,d), 6.2-11.2 (b,e) and 8.7-14.5~Mm (c,f) extracted from the MHD wave-field simulations of $\S$\ref{MHD.hanasoge}. Light solid lines represent Gaussian frequency filtering centered 3.5 mHz, dashed lines represent 4.0~mHz and bold solid lines represent 5.0~mHz.}
\label{fig:dtmaps}
\end{figure}

The simulated Doppler velocity data-cube of $\S$\ref{MHD.hanasoge} (extracted at an observational height of 200~km above the photosphere) had dimensions of $200\times200$~Mm$^2$ $\times$ $512$ minutes, with a spatial resolution of $0.781$~Mm and a cadence of 1 minute. For the time-distance calculations, we compute cross covariances of oscillation signals at pairs of points on the photosphere (source at $\bf{r_1}$, receiver at $\bf{r_2}$) based on a single-skip center-to-annulus geometry (see e.g. \citealp{cbk2006}). We cross correlate the signal at a central point with signals averaged over an annulus of radius $\Delta=|\bf{r_2}-\bf{r_1}|$ around that center. Firstly, we filter out the $f$-mode ridge. Subsequently, standard Gaussian phase-speed in conjunction with Gaussian frequency filters centered at 3.5, 4.0 and 5.0~mHz with 0.5 mHz band-widths are applied in order to study frequency dependencies of travel times (e.g. \citealp{bb2006}; \citealp{cr2007}). The annular sizes and phase-speed filter parameters used in estimating the times shown in Figures~\ref{fig:dtmaps} and~\ref{fig:azim} (including the central phase speed ($v$) and full width at half-maximum (FWHM) used) are outlined in Table \ref{tab:filters}. The point-to-annulus cross-covariances are fitted by two Gabor wavelets (e.g., \citealp{cbk2006}) to extract the required travel times.  

In order to compare theory with simulation, we estimate centre-to-annulus mean time shifts, $\delta\tau_{mean}$, using the MHD ray tracing technique of $\S$\ref{MHD.moradi} for the same sunspot model ($\S$\ref{sunspot}). The single-skip magneto-acoustic rays do not require filtering. Instead, they are propagated from the upper turning point of their trajectories, in both the positive and negative $x$ directions, at a prescribed frequency with horizontal increments of 1~Mm across the sunspot. The required range of horizontal skip distances are obtained by altering the shooting angle at which the rays are initiated. The skip distances are then binned according to their travel path lengths, $\Delta$, while the travel times are averaged across both the positive and negative horizontal directions. For both forward models (i.e., $\S$\ref{MHD.hanasoge}, \ref{MHD.moradi}), we only concern ourselves with the mean \emph{phase} time shifts. 

\section{RESULTS AND ANALYSES}
\subsection{Travel-Time Profiles I. MHD Wave-Field Simulations}\label{mhdtt}
Figure \ref{fig:dtmaps} shows maps of $\delta\tau_{mean}$ as well as the frequency filtered azimuthal averages of $\delta\tau_{mean}$ obtained using time-distance center-to-annulus measurements for the measurement geometries indicated in Table \ref{tab:filters}. The $\delta\tau_{mean}$ map for $\Delta=6.2-11.2$~Mm clearly displays positive travel-time shifts, reaching a maximum of around 25 seconds at spot center.  A similar travel-time shift is observed from the azimuthal average of $\delta\tau_{mean}$ when a frequency filter centered at 3.5~mHz is applied to the data.  We also observe the magnitude of the positive $\delta\tau_{mean}$ steadily decrease as we increase the frequency filter to 4.0~mHz, with negative $\delta\tau_{mean}$ starting to appear in the profile, and by 5.0~mHz the travel times observed inside the sunspot are completely negative. For the larger annuli, negative time shifts of increasing magnitude are consistently observed as we increase the central frequency of the filter. In fact, all $\delta\tau_{mean}$ maps for $\Delta$ larger than $8.7-14.5$~Mm that we measured displayed similar $\delta\tau_{mean}$ behavior to the $6.2-11.2$ and $8.7-14.5$~Mm bins (albeit with smaller time shifts). 

It is important to take note of both the signs of the travel-time perturbations and their apparent frequency dependence. Positive $\delta\tau_{mean}$ have traditionally been interpreted as indicative of a region of slower wave propagation in the shallow subsurface layers beneath the spot, while negative times are of a wave-speed enhancement. So in essence, the $\delta\tau_{mean}$ profiles that we have derived from the simulation would appear to indicate a traditional ``two-layered" wave-speed structure (e.g., \citealp{kds2000}; \citealp{cbk2006}) beneath the sunspot. However as can be seen in Figure~\ref{fig:mhd}, the thermal profile of our model atmosphere is a ``one-layer'' sunspot model ($\delta c^2/c^2 < 0$) and of the order of $\sim-40\%$. Similarly, changes in the sub-surface wave speed, $(c^2 + a^2)/c_0^2 -1$ (where $c_0$ is the unperturbed sound speed), lie only in the positives $\sim$ 0--250\% (not shown here), with the greatest enhancements seen near the surface. The large decrease in the sound speed we observe in our model also raises the possibility that current methods of linear inversion may lie beyond their domains of applicability. 

We also observed that travel times associated with the smallest measurement geometry are most sensitive to the phase-speed filter used, i.e., when the phase-speed parameters were adjusted to filter all background power below the $p_1$ ridge, negative $\delta\tau_{mean}$ were obtained. This behavior was noted by \cite{bb2008}, who determined the causative factor to be the background power between the $p_1$ and $f$ ridges. It is unsettling that the sign of the time shift may be reversed at will, through small changes in the filter width and center.

\subsection{Travel-Time Profiles II. MHD Ray-Path Simulations}\label{raytt}	
\begin{figure}[ht]
\begin{center} 
\epsscale{1.0}
\plotone{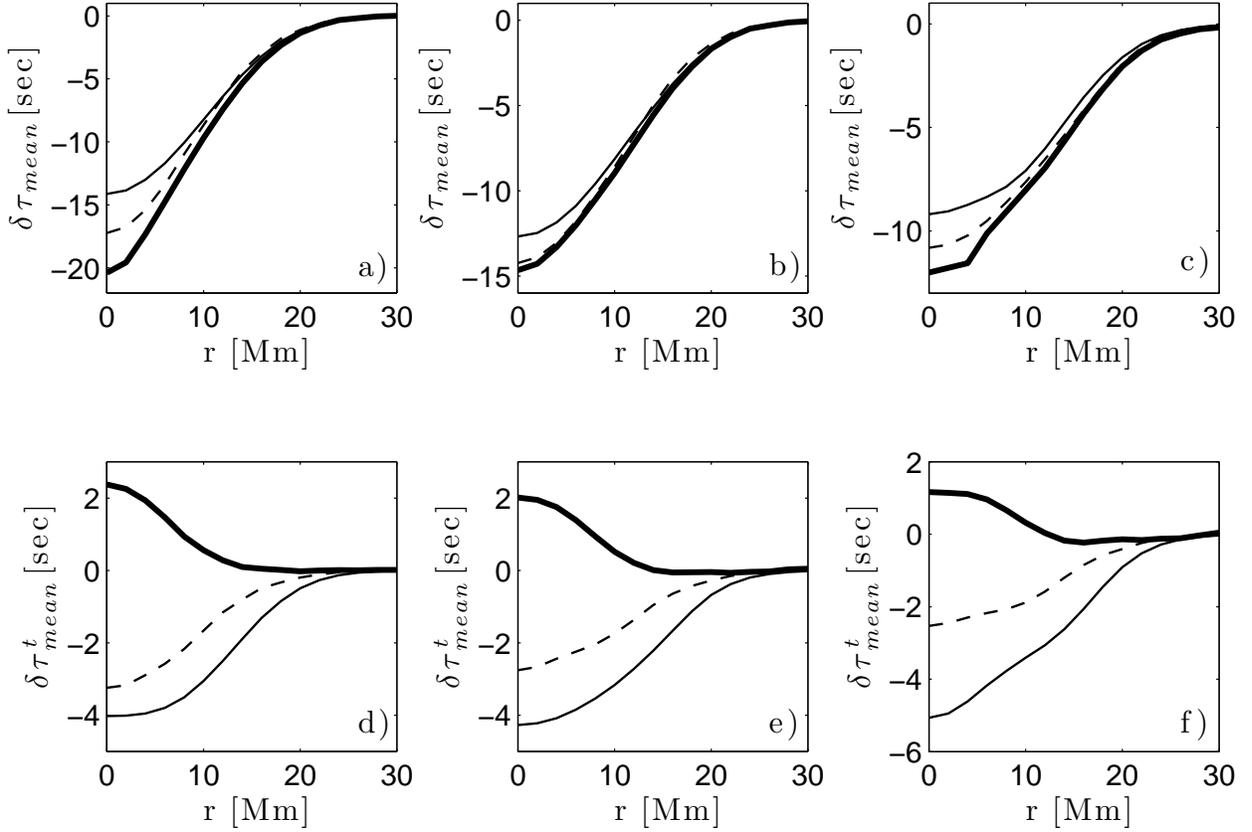}
\end{center}
\caption{Center-to-annulus ray $\delta\tau_{mean}$ and  $\delta\tau^t_{mean}$ profiles for $\Delta$= 3.7-8.7 (a,d), 6.2-11.2 (b,e), and 8.7-14.5~Mm (c,f) computed using the MHD ray calculation recipe of $\S$\ref{MHD.moradi}. Light solid lines represent 3.5 mHz, dashed lines represent 4.0~mHz and bold solid lines represent 5.0~mHz.}
\label{fig:azim}
\end{figure}

Figure \ref{fig:azim} (frames a-c) show the resultant $\delta\tau_{mean}$ profiles derived from the MHD ray tracer for identical measurement geometries as used for the time-distance calculations. The similarities between between the ray $\delta\tau_{mean}$ profiles and their time-distance counterparts in Figure \ref{fig:dtmaps} are striking. Firstly, the ray travel-time perturbation profiles contain predominantly negative travel-time shifts for all frequencies, albeit with slightly smaller magnitudes. Secondly, a similar frequency dependence of $\delta\tau_{mean}$ is also observed. Generally, high frequency rays propagated within the confines of a magnetic field are expected to i) travel faster and ii) propagate longer distances than low frequency rays \citep{cally2007,mc2008}. However, one significant difference we can observe in these profiles is the absence of any positive travel-time shifts for the $\Delta=3.7-8.7$~Mm bin. This is significant because the exclusively negative $\delta\tau_{mean}$ we observe across all geometries not only reflects the one-layered wave- and sound-speed profiles below the surface, but also highlights the effects that phase-speed filtering can have on time-distance measurements (recall that ray calculations require no such filtering). Nonetheless, the overall self-consistency between these results and those in $\S$\ref{mhdtt} are very encouraging, despite the 2D nature of the ray calculations.

Given the fact that ray theory appears to succeed in capturing the essence of the travel-time variations as derived from the MHD simulations, we can isolate the thermal component of the measured  $\delta\tau_{mean}$ using the same approach as \cite{mc2008} to ascertain the contribution to the travel-time shifts from the underlying thermal structure. To do this, we essentially re-calculate the ray paths in the absence of the flux tube while maintaining the modified sound-speed profile obtained in $\S$\ref{sunspot}. The resulting \textit{thermal} travel-time perturbations, $\delta\tau^t_{mean}$, would then be purely a result of thermal (sound-speed) variations along the ray path. 

The resulting $\delta\tau^t_{mean}$ profiles, presented in Figure \ref{fig:azim} (frames d-f), surprisingly show that, even without the magnetic field, ray theory produces negative travel times -- the exception being for rays propagated at 5.0~mHz. This indicates that the contribution from the underlying thermal structure is significant enough to modify the upper turning point of the ray paths, thus shortening the ray travel times. The appearance of negative travel times for a model with a decrease in sound speed would appear to be somewhat counterintuitive, since from standard ray theory, one would expect negative time shifts with increases in sound speed. The most likely explanation for this phenomenon, is that since both the sound speed and density differ from the quiet Sun, consequent changes in the acoustic cut-off frequency ($c/2H$, where $H$ is the density scale height) in the near-surface regions of our model modifies the the ray path for waves with frequencies less than 5.0~mHz quite significantly, thereby causing negative travel-time shifts. 

However, when comparing with the time perturbations derived from calculations that include the magnetic field (i.e. Figure \ref{fig:azim} frames a-c), MHD effects appear to be dominant contributors to the observed time shifts. This is perhaps most evident for $\Delta= 3.7-8.7$~Mm (Figure~\ref{fig:azim} d), where the thermal contribution at spot center appears to make up approximately 11\% of $\delta\tau_{mean}$ (i.e. Figure \ref{fig:azim} a) at 5.0~mHz, 18\% at 4.0~mHz and 28\% at 3.5~mHz. For the largest bin, we see a similar contribution at 4.0 and 5.0 mHz, but a much greater contribution at 3.5~mHz (45\% of $\delta\tau_{mean}$).  


\section{SUMMARY AND DISCUSSION}
Incorporating the full MHD physics into the various forward models used in local helioseismology is essential for testing inferences made in regions of strong magnetic fields. By comparing numerical simulations of MHD wave-field and ray propagation in a model sunspot, we find that: i) the observed travel-time shifts in the vicinity of sunspots are strongly determined by MHD physics, although sub-surface thermal variations also appear to affect ray timings by modifying the acoustic cut-off frequency, ii) the time-distance travel-time shifts are strongly dependent on frequency, phase speed filter parameters and the background power below the $p_1$ ridge, and finally iii) MHD ray theory succeeds in capturing the essence of center-to-annulus travel-time variations as derived from the MHD simulations. 

The most unsettling aspect about this analysis is that despite using a background stratification that differs substantially from Model S \nocite{cdetal1996} (Christensen-Dalsgaard et al. 1996) and a flux tube that clearly lacks a penumbra, the time shifts still look remarkably similar (at least qualitatively) to observational time-distance analyses of sunspots. Preliminary tests conducted with different sunspot models (e.g.,  different field configurations, peak field strengths etc.) have also provided similar results. Given the self-consistency of these results, as derived from both forward models, it could imply that we are pushing current techniques of local helioseismology to their very limits. It would appear that accurate inferences of the internal constitution of sunspots await a clever combination of forward modeling, observations, and a further development of techniques of statistical wave-field analysis.

%

%
\clearpage

\begin{deluxetable}{rrr}
\tablecolumns{3}
\tablewidth{0pc}
\tablecaption{Annuli Radii and Phase-Speed Parameters}
\tablehead{
\colhead{$\Delta$ (Mm)} & \colhead{$v$ (km/s)} & \colhead{FWHM (km/s)}  }
\startdata
3.7--8.7 & 17.71 & 11.94 \\
6.2--11.2 & 21.11  & 11.94\\
8.7--14.5 & 24.36 & 11.94 
\enddata
\label{tab:filters}
\end{deluxetable}

\clearpage

\end{document}